\documentclass[useAMS,usenatbib]{mn2e}
\usepackage{graphicx}
\title[V1481 Ori: SB2 with accreting component]{
Physical parameters and long-term photometric variability of V1481 Ori, a SB2 member of Orion Nebula Cluster
with an accreting component.}
\author[S. Messina et al.]{ S. Messina$^{1}$\thanks{E-mail:
sergio.messina@oact.inaf.it}, P. Parihar$^{2}$, K. Biazzo$^{1}$, 
A.F. Lanza$^{1}$, E. Distefano$^{1}$, C.H.F. Melo$^{3}$, 
\newauthor
D.H. Bradstreet$^{4}$ and W. Herbst$^{5}$\footnotemark[1]\thanks{Based on data collected at the Indian Institute of Astrophysics (IIA) and European Southern Observatory (ESO).}\\
$^{1}$INAF-Catania Astrophysical Observatory, via S.Sofia 78, I-95123 Catania, Italy\\
$^{2}$Indian Institute of Astrophysics, Bangalore 560034, India\\
$^{3}$ESO - European Southern Observatory, Alonso de Cordova 3107, Vitacura Casilla 19001, Santiago 19, Chile\\
$^{4}$Eastern University, St. Davids, PA, USA\\
$^{5}$Astronomy Department, Wesleyan University, Middletown, CT, 06459, USA}
\begin{document}
\date{}
\pagerange{\pageref{firstpage}--\pageref{lastpage}} \pubyear{}
\maketitle
\label{firstpage}
\begin{abstract}
We present the results of our analysis on V1481 Ori (JW\,239), a young SB2 in the Orion Nebula Cluster
with a circumbinary disc accreting on the lower-mass component.
The analysis is based on high-resolution spectroscopic data and high-quality photometric time series about 20-yr long. 
Thanks to the spectroscopy, we confirm the binary
nature of this system consisting of M3 + M4 components and  derive the mass ratio M$_{\rm B}$/M$_{\rm A}$ = 0.54, 
a variable luminosity ratio L$_{\rm B}$/L$_{\rm A}$ = 0.68--0.94, and an orbital period P$_{\rm orb}$ = 4.433\,d. The photometric data allowed us to measure the rotation
periods of the two components P$_{\rm phot}$ = 4.4351\,d  and they are found to be synchronized with the orbital period. The simultaneous
modeling of V-, I-band, and radial velocity curves in the 2005 season suggests that the variability is dominated by one
hot spot on the secondary component covering at least $\sim$3.5\% of the stellar surface and about 420\,K hotter
than the unperturbed photosphere. Such a spot may originate from the material of the circumbinary disc accreting onto the 
secondary component.  We also detect an apparent 6-yr periodic variation in the position of this hot spot, which is inferred from the phase migration of the light curve maximum,
which we interpret as due to either the presence of surface differential rotation as large as 0.065\%, a value compatible with the fully convective components, or
to a periodic exchange of angular momentum between the disc and the star, which implies a minimum magnetic field strength of 650\,G at the stellar surface.

 \end{abstract}


\section{Introduction}
Close binary systems with low-mass components in star-forming regions are valuable astrophysical laboratories
to address a number of important issues such as accretion, differential rotation, and rotation synchronization.
Current models of accretion processes make different predictions (see, e.g., \citealt{Artymowicz96}; \citealt{de Val-Borro11}) 
about the component(s) on which the accretion takes place, especially in the case the components have different masses.
A comparison of model results with real cases is, therefore, highly useful.\\
Another important issue concerns the presence of differential rotation. The components of the system are fully convective yet, and, 
on theoretical grounds, the rotation state  is expected to be very close to a rigid-body state (see, e.g., \citealt{Kuker97}). 
As a consequence, the stellar magnetic field  is likely to be generated by a distributed dynamo like the $\alpha^2$ dynamo, rather than an $\alpha\Omega$ dynamo. 
The field geometry is expected to be non-axisymmetric and to show no cyclical behavior. 
Moreover,  \cite{Kuker01}, modeling the differential rotation of the Sun, derived its PMS evolution, and found that the total shear $\delta\Omega$ on the Hayashi track is small and it depends on spectral type rather than on rotation rate. 
 The lower the effective temperature the more rigid the rotation (see, e.g., \citealt{Barnes05}; \citealt{Collier07}). \rm  These expectations are supported by our own analysis as well as by other earlier \rm observational studies 
(e.g., \citealt{Johns-Krull96}; \citealt{Rice96}) that did not detect any significant evidence for  surface differential rotation (SDR) in 
classical T Tauri (CTTS) stars. For example, \cite{Cohen04},  basing their analysis on an approximately 4 yr long time series data of a sample of IC 348
members ($\sim$ 2--3 Myr),  neither detect clear evidence of activity cycles nor of photometric rotation period variations attributed to differential rotation. 
However,  persistent  spots that do not migrate \rm may be responsible for the SDR non-detection.\\
Another interesting issue concerns the time scale for the synchronization of the rotation of the system's components with the orbital period. To address this, we require measurements of photometric rotation periods of both components and of the orbital period. \\
A few such interesting close binary systems have been monitored spectroscopically and photometrically  over a long time base  as a part of a program 
focussed on Young Stellar Objects (YSO) in the Orion Nebula Cluster (ONC) begun by Parihar and collaborators in 2004 (\citealt{Parihar09}),
using the facilities available at the Indian Astronomical Observatory (IAO).
A number of the ONC members monitored at IAO have been also photometrically monitored at the Van Vleck Observatory (VVO) on the campus of Wesleyan University since 1991 by Herbst and collaborators (see, e.g., \citealt{Herbst00}). These two monitoring programs have  provided us with a photometric data time series  almost 20 year long for about one hundred  ONC members. The data collected over such a long  time span, together with spectroscopic data,   allow us to explore the accretion processes, presence of SDR, and rotation/orbital period synchronization.\\
In the present paper, we report the results obtained for one interesting ONC cluster member - the spectroscopic binary V1481 Ori. We find that this system exhibits periodic photometric variations that may originate from the presence of an accretion hot spot on the photosphere of the secondary component. The hot spot is carried in and out of view by the stellar rotation and produces periodic flux enhancements. Although the system does not exhibit any evidence of photometric rotation period variation, unexpectedly, it shows a periodic variation of the phase of the light curve maximum.  This oscillation of the phase of maximum may arise from a variation of the angular velocity due to a periodic latitude migration of the hot spot on a differentially rotating star. An alternative explanation is that the disc and the star exchange angular momentum back and forth in a cyclic fashion.\\
The literature on V1481 Ori is reviewd in Sect.\,2, the spectroscopic and photometric observations are presented in Sect.\,3, the photometric analysis is in Sect.\,4, and the derived orbital and physical parameters are given in Sect.\,5. Discussion and conclusions are presented in Sect.\,6 and 7.

\section{V1481 Ori}
V1481 Ori ($\alpha$ = 05:35:03.92, $\delta$ = $-$05:29:03.35, J2000.0; V = 15.45\,mag,  V$-$I = 2.40\,mag) is a low-mass member of the Orion Nebula Cluster. 
First catalogued with sequential number 1725 by \cite{Parenago54} in his survey of stars in the Orion Nebula area, it was subsequently discovered to be variable by \cite{Rosino56} from analysis of photographic plates. \cite{Jones88}, who assigned the number  JW 239,  also found evidence of variability.  The rotation period P = 4.46\,d was first measured by \cite{Edwards93},  after this star was included in the photometric monitoring program at VVO in 1991 (\citealt{Attridge92}). It is designated as V1481 Ori in the 76th Name-List of Variable Stars (\citealt{Kazarovets01}). \\
Basic stellar parameters were first determined by \cite{Hillenbrand97} who estimated  a probability of membership to ONC of 99\%,  mass M = 0.24\,M$_{\odot}$, radius R = 2.73\,R$_{\odot}$, effective temperature T$_{\rm eff}$ = 3590\,K, interstellar absorption A$_{\rm V}$ = 0.47 mag, and spectral type M1.5, whereas \cite{Edwards93} assigned a later $\sim$M4 spectral type.  \cite{Da Rio10} have estimated a smaller reddening A$_{\rm V}$ = 0.33 mag and a larger mass M $\simeq$ 0.41\,M$_{\odot}$. However, all these measurements referred to the unresolved system. Recently, V1481 Ori was discovered  to be a SB2 system by \cite{Tobin09}. \cite{biazzoetal2009} found additional evidence of its binary nature from a cross correlation function analysis, and derived the following values for the projected rotational velocities of the two components: $v\sin{i}_A$ = 19.2$\pm$1.1\,km\,s$^{-1}$ and $v\sin{i}_B$ = 16.1$\pm$0.9\,km\,s$^{-1}$. 
V1481 Ori exhibits variable  H$\alpha$ line emission: EW = $-$13.0\,\AA\,(in March 2004; \citealt{Sicilia05}), +0.44\,\AA\,(in Jan 2005; \citealt{Da Rio09}), $-$12.75\,\AA\,(in Feb 2007; \citealt{Parihar09}),  and $-$17.7\,\AA\, (\citealt{Furesz08}).\\ \rm
The possible presence of a disc is indicated by near infrared (NIR) excess in both 2MASS and IRAC/Spitzer photometry (\citealt{Werner04}). \cite{Rebull06} measure a value of [3.6] - [8] = 1.3\,mag, indicative of an active accretion disc, and found hint that the accretion was variable. \rm Based on these results, they classified it as CTTS. 
On the other hand, other measurements do not show any excess, like the values of  $\Delta$[I$-$K] = $-$0.03 mag measured by \cite{Rodriguez10}, or $\Delta$[I$-$K] =  0.38 mag by \cite{Hillenbrand98}, which is a value very close to the boundary adopted to distinguish disc from disc-less stars. The more recent WISE (Wide-field Infrared Survey Explorer) observations (\citealt{Cutri13})  demonstrate a prominent IR excess, leaving no doubt about the presence of an accretion disc.\\
V1481 Ori is also an X-ray source, first detected by the ROSAT satellite as part of surveys of ONC (\citealt{Gagne95}, \citealt{Geier95}), and subsequently by the CHANDRA satellite (\citealt{Feigelson02}, \citealt{Getman05}).

\section{Observations}
\subsection{Spectroscopy}

The spectroscopic observations of V1481\,Ori were obtained in Period 76A at the ESO/VLT telescope (Chile). Spectra were collected within the FLAMES (Fibre Large Array Multi Element Spectrograph)  GTO  with the multi-object 
GIRAFFE spectrograph in MEDUSA mode\footnote{This is the observing mode in {\sc flames} in which 132 fibres with a projected 
diameter on the sky of 1\farcs{2} feed the GIRAFFE spectrograph. Some fibres are set on the target stars and others on the sky 
background.}. One of these spectra was analyzed by \cite{biazzoetal2009} with the aim to study the disc-locking scenario 
for low-mass stars in the Orion Nebula Cluster. Observations were carried out in 2005-2006 from October 15th to January 20th 
($\sim$8.4 h on 11 nights). The HR15 GIRAFFE setup ($R \sim 19\,300$, $\lambda=659.9-695.5$\,nm) was chosen. The log 
of spectroscopic observations is given in Table~\ref{tab:observations}. A total of eleven spectra of V1481\,Ori were obtained 
and reduced using the GIRAFFE Base-Line Data Reduction Software (girBLDRS; \citealt{blecha2000}). Sky subtraction was 
applied following the prescriptions reported in  \cite{biazzoetal2009}.

\begin{table} 
\caption{Log of spectroscopic observations.}
\label{tab:observations}
\begin{center}  
\begin{tabular}{ccc}
\hline
\hline
\noalign{\smallskip}
Date      &  UT      & $t_{\rm exp}$ \\
(d/m/y)    & (h:m:s)  &  (s)    \\ 
\noalign{\smallskip}
\hline
\noalign{\smallskip}
15/10/2005 & 07:34:35.6 & 2820  \\ 
16/10/2005 & 07:21:02.3 & 2820  \\ 
17/10/2005 & 07:23:30.4 & 2820  \\ 
18/10/2005 & 06:48:58.4 & 2054  \\ 
19/10/2005 & 07:23:01.9 & 2820  \\ 
20/10/2005 & 07:17:21.2 & 2820  \\ 
21/10/2005 & 06:43:17.8 & 2820  \\ 
04/11/2005 & 06:25:09.4 & 2820  \\ 
05/11/2005 & 05:05:07.5 & 2820  \\ 
18/01/2006 & 03:28:52.9 & 2820  \\ 
20/01/2006 & 04:05:21.6 & 2820  \\ 
\noalign{\smallskip}
\hline
\end{tabular}
\end{center}
\end{table}



\subsection{Photometry}
The photometric analysis is based on observations collected during two in-progress monitoring programs of ONC members. The first and longest one is the program carried out by Herbst and collaborators at the Van Vleck Observatory since 1991 (VVO; see, e.g., \citealt{Herbst00}).  The second and more recent one is the observing program carried out by Parihar and collaborators at the Indian Astronomical Observatory  since 2003 (IAO;  see \citealt{Parihar09}).

In order to get a homogeneous time series of I-band magnitudes, we proceeded as follows. 
We selected first the time intervals during which both IAO and VVO photometry were almost contemporary (in the years 2004 and 2005). 
Then, we folded the light curves with the already known rotation period (from \citealt{Herbst00}) and, finally, added a magnitude offset  to the VVO data to match with the IAO light curves with the aim to minimize the magnitude dispersion versus rotation phase. This offset  $\Delta$I = 0.04\,mag arises from either difference between the Johnson-Cousin filter at VVO and Bessel filter at IAO and the uncertainty on the magnitude standardization. The same magnitude offset was applied to all VVO data.\\
In December 1994, V1481 Ori was simultaneously observed at VVO and by \citealt{Stassun99} with the 1-m telescope at the Wise Observatory, as part of an independent program. As previously done, we added an offset to the differential magnitudes by  \citealt{Stassun99} in order to minimize the magnitude dispersion versus rotation phase with respect to the VVO data.
In the end, the precision with which photometric sets coming from different telescopes were combined together was comparable to the  average photometric precision of each independent data set.

We applied a 3$\sigma$ clipping filter to remove outliers on each seasonal time series,  where $\sigma$ is the standard deviation of the seasonal magnitude time series. \rm In principle, some outlier may represent a flare event. However,
for the purposes of the present study we focused on  only variability arising from the rotational modulation. Then, we averaged observations collected within a time interval of 1 hr. 
More details on the data reduction are given in \cite{Parihar09} and \cite{Herbst00}.\\
Our final dataset consists of 1552 measurements spanning a time interval of about  20 yr from October 1991 to March 2011. A summary of the observations is given in Table\,\ref{data}.
 The complete photometric time series (HJD versus I magnitude) is made available 
as online Table. \rm
In the years from 2004 to 2011 we obtained contemporary V and I magnitude time series at IAO, and also R magnitudes during the 2008 season. 

\section{Photometric analysis}
\subsection{Correlation analysis}

\begin{figure}
\begin{minipage}{10cm}
\centering
\includegraphics[scale = 0.25, trim = 0 0 0 0, clip, angle=0]{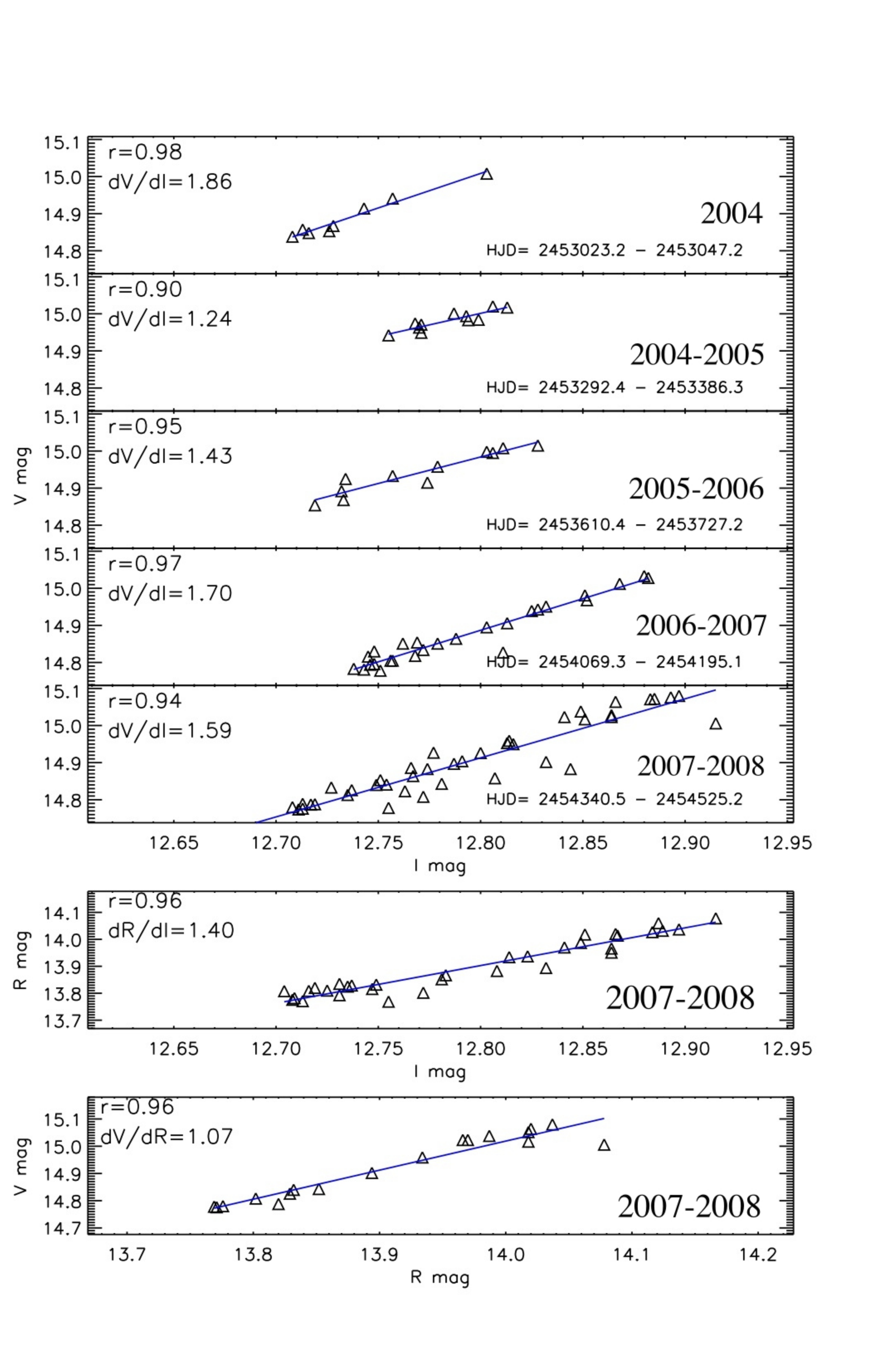}
\end{minipage}
\vspace{-0cm}
\caption{\label{correl} Correlation analysis of V versus I magnitudes for the observation seasons from 2004 to 2008, and among I, R, and V magnitude for the 2008 season. 
The correlation coefficients, the slopes of the linear fits, and the date range are plotted as labels.}
\end{figure}

The multi-band observations collected between 2004 and 2008 allow us to infer some additional information on the nature of the observed variability. 
For each seasonal light curve we have carried out correlation and linear regression analyses between colors and magnitude variations,  which we measured in our light and color curves,
by computing  correlation coefficients  (r),  their  significance level  ($\alpha$), and slopes of linear fits (Fig.\,\ref{correl}).
The significance level $\alpha$ represents the probability of observing a value
of the correlation coefficient larger than r for a random sample
having the same  number of observations and degrees of freedom (\citealt{Bevington69}). 

We find that the Pearson linear correlation coefficients (r) are larger than 0.90 in every season and their significance levels, $\alpha$, are larger than 99\%. 
The slopes of the V versus I variations, derived from linear regression analysis, vary from season to season in the range 
 dV/dI = 1.24--1.86 (dR/dI = 1.40, dV/dR = 1.07 in 2008). 
 Similarly, we computed the slopes of the V versus V$-$I variations and found them to vary in the range dV/d(V$-$I) = 1.84--2.04, 
 with a mean value of 1.89.\\
 Correlation analysis allows us to investigate the origin of magnitude and color variations. For example,
if these variations originate from a single spot or group of small
spots, as well as if they originate from two different, but spatially and temporally correlated types of inhomogeneities, e.g.
cool spots and hot faculae, we expect these quantities to be correlated. On the contrary, a poor correlation or its absence will tell
us that magnitude and colors are affected by at least two mechanisms, which are operating independently from each other, either
spatially or temporally. The regression analysis is also important to infer
information on the properties of photospheric inhomogeneities,
since their averaged temperature mostly determines the slope of
the trends. Indeed, surface inhomogeneities with different areas but
constant temperature will determine magnitude and color variation of different amplitude but with a rather constant ratio. \\
Assuming that  magnetic activity is present in both components of V1149 Ori, this circumstance may also play a role in decreasing the expected 
correlation, since the variabilities arising from the two components are not necessarily correlated. 
However, we find that the light variations in all bands are strongly correlated,  which indicates that either the photometric variability of one component likely dominates
over the variability from the other component or they have same periodicities. 
The spectroscopic analysis in Sect.\,5 will show that the secondary component dominates the observed variability, which arises
from a hot spot produced by the infalling gas from the accreting disc. Therefore, in the subsequent analysis we will neglect the
contribution to variability by the primary component that will be assumed to be negligible.

The brightest magnitudes and bluest colors of V1481 Ori were observed in the 2010/2011 season and they are V = 14.62$\pm$0.01\,mag, I = 12.61$\pm$0.01\,mag, and V$-$I = 2.01$\pm$0.02\,mag, whereas the faintest and reddest were observed in 1994/1995 and they are V = 15.45$\pm$0.01\,mag, I = 13.05$\pm$0.01\,mag, and V$-$I = 2.40$\pm$0.02\,mag.
Therefore, the stars exhibit a variability amplitude of $\Delta$V = 0.83$\pm$0.015\,mag,  $\Delta$I = 0.44$\pm$0.015\,mag, and  $\Delta$(V$-$I) = 0.39$\pm$0.03\,mag.
When modeling the observed magnitude variations, we must keep in mind that they refer to the unresolved system. Therefore,
owing to a dilution effect due to the presence of the primary component, which in the present case is a factor 2.18 derived from the magnitude difference between the two components in the I band, the total magnitude variations of the secondary component are larger by about the same factor
than those observed.


\subsection{Search for rotation period}
The light curves of low-mass active stars are known to be characterized by changes of amplitude and shape
over different timescales. During the PMS evolution, WTTS have generally quite stable light curves that maintain similar amplitude and shape for several rotation cycles. On the contrary, CTTS can exhibit significant variations during the same season or from one observation season to the subsequent one,  owing to variable accretion phenomena (see, e.g., \citealt{Grankin07},\,\citealt{Grankin08}). To search for such variations, we decided to divide the whole time series into a number of segments, corresponding to single observation seasons. Therefore, our analysis was carried out on 19 different light curve segments as well as on the complete time series (see Table\,\ref{data}).

The period search was carried out using the Lomb-Scargle periodogram (\citealt{Scargle82}). The false alarm probability (FAP), that is the probability that a periodicity is not true but simply arises from  noise associated to the data, was computed using the bootstrap approach proposed by Herbst et al. (2000), i.e., by generating 1000 artificial light curves obtained from the real one, keeping the date but scrambling the magnitude values. We performed the periodogram analysis on each fake randomized dataset and determined the power level corresponding to a false alarm probability FAP = 0.01. We decide to consider only periodicity with  FAP $<$ 0.01, that is with confidence level larger than 99\%.

In Table\,\ref{period} we summarize the results of our period search. The same periodicity, which we attribute to the  rotation period, was detected with high confidence level in 18 of 19 light curves, as well as in the complete time series. The uncertainty on the period determination was computed following \cite{Lamm04}. In Fig.\,\ref{lightcurve} we display the seasonal time series versus HJD (left column); the periodogram with over-plotted (horizontal dashed line) and the power level corresponding to FAP = 0.01 (middle column); the I-band light curves phased with the ephemeris HJD = 2453658.8320 + 4.4351$\times$E (right column). We note that in only one season the period is not detected with FAP $<$ 0.01.  Nonetheless, the phased light curve is very smooth and it allows us to accurately determine the phase of maximum.
The average rotation period is P = 4.439 $\pm$ 0.011\,d. 
The largest period variation in the series listed in Table\,\ref{period} is 0.04d, which is smaller than the average uncertainty on the period determination (0.05d). Considering the average period and its uncertainty, all seasonal period determinations differ from the average by less than 2$\sigma$.
Therefore, we infer that our star does not show any rotation period variation larger than 0.7\%, which is the minimum variation at 3$\sigma$ level detectable by  our analysis. \\
When we plotted the series of 19 light curves using the average rotation period to compute the rotation phases, we noted a linear migration of the phases of light curve maximum (owing to the presence of  hot spots, we consider the light maximum). We made use of the  O$-$C diagram to find the value for which the series of maxima did not exhibit any linear trend (see bottom panel of Fig.\,\ref{result}), and found the following rotation period P = 4.4351\,d. In the right column of Fig.\,\ref{lightcurve},  lightcurves are phased with the ephemeris HJD = 2453658.8320  
+ 4.4351$\times$E. This is the rotation period of the component hosting the hot spot.  As we will see from the light curve modeling and the spectroscopic analysis, the hot spot is hosted by the secondary component. Since the primary component is brighter than the secondary component, its photometric variability is expected to be significant even if the system is not resolved. Now, if the primary component would rotate with a period different that the rotation period of the secondary component, we would expect to find its modulation in the photometric timeseries. However, we note that in neither in the periodogram of the complete photometric series nor in the seasonal periodograms we found evidence for a significant secondary period. 
Therefore, we can consider that both components have
the same rotation that is synchronized with the orbital period. This is particularly interesting because it proves that the components of a close binary can be synchronized during the PMS stage of their evolution. However, we cannot completely rule out the possibility that the primary has a different rotation period, but being quite inactive, although this circumstance is quite unlikely, we do not detect it neither in the light curves nor in the periodograms. \rm


\subsection{Pooled variance analysis}

We applied also the pooled variance analysis as proposed by \cite{Dobson90}, \cite{Donahue92}, and \cite{Donahue97} to our data time series. The pooled variance is the average variance of the data subsets of length $\tau$. The variance profile plotted in Fig.\,\ref{pool} allows us to identify three characteristic time intervals. The rotational modulation of the flux arising from the presence of  active regions produces the first rise of the variance, up to a timescale about 5 days. The pooled variance stays more or less constant for a time interval from about 5 days up to about 1000 days, that should correspond to the lifetime of the largest surface brightness inhomogeneities. Our photometric times series is not long enough to infer additional information for timescales longer than 1000 days.
Therefore, we confirms that the bright region responsible for the observed variability (which we will show to be the hot spot on the accreting secondary component) is quite stable for about 3 years, which makes the season-to-season variation of the light curve very smooth.

\subsection{Search for long-term cycles}
The Lomb-Scargle periodogram was also used to investigate the presence of significant long-term periodicities. We computed the periodograms of four different variability indicators:  I-band magnitudes,  brightest and faintest \rm light curve magnitudes for each light curve segment,  light curve amplitudes, and  phases of light curve maximum. 

\begin{figure*}
\begin{minipage}{18cm}
\centering
\includegraphics[scale = 0.7, trim = 0 0 0 80, clip, angle=90]{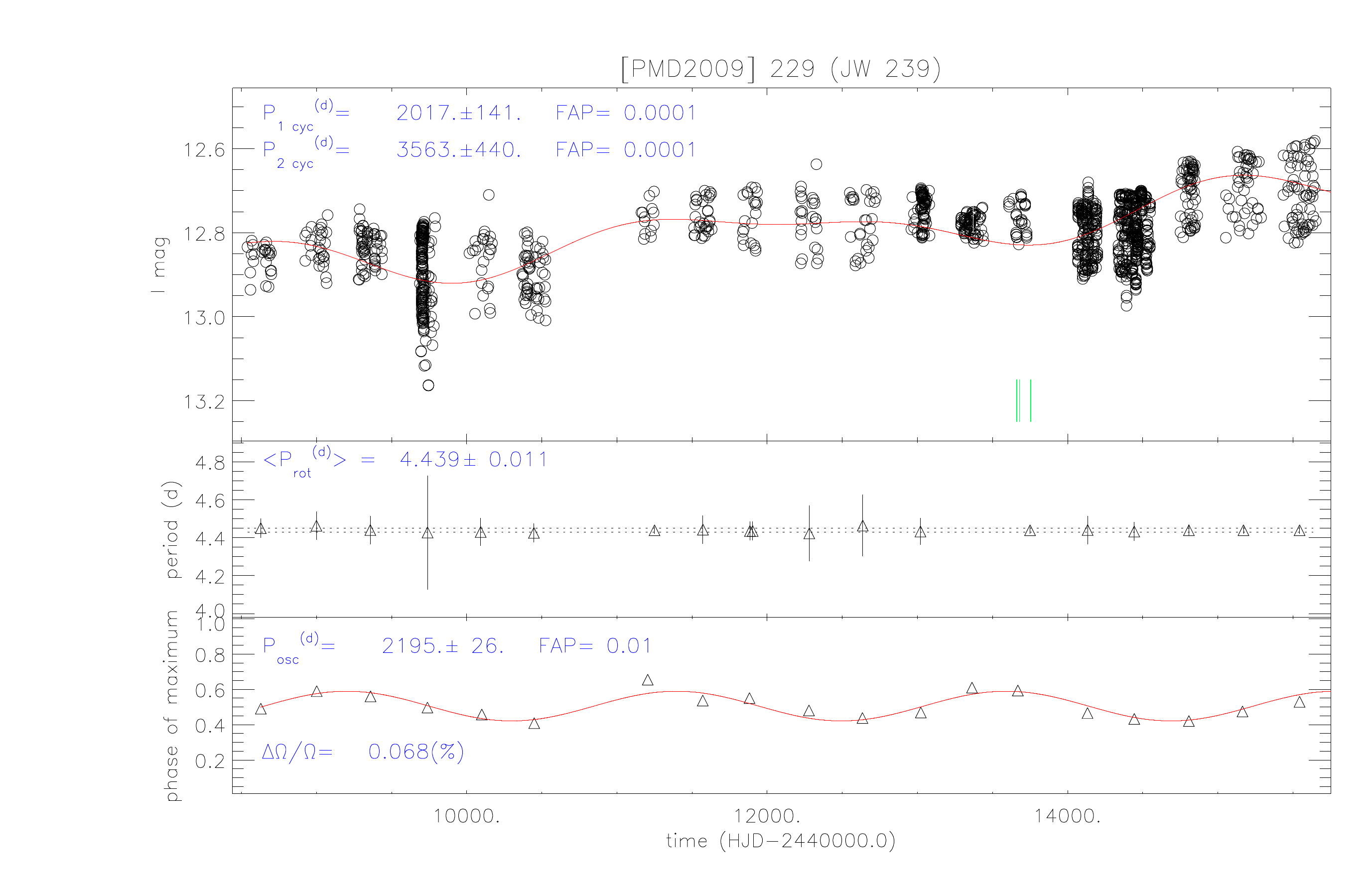}
\end{minipage}
\vspace{-0cm}
\caption{\label{result} \it Top panel\rm: I magnitude time series of V\,1149 Ori versus time. The solid line represents the sinusoidal fit with the cycle periods of P$_1$=2046d and P$_2$=3470d. Dotted green vertical bars indicate the epochs of spectroscopic observations. \it Middle panel: \rm 
Time sequence of the seasonal rotation periods listed in Table\,\ref{result}. Dotted lines indicate the 1-$\sigma$ uncertainty on the average period. \it Bottom panel: \rm Phase of the light curve maximum versus time. Solid line represent the sinusoidal fit with the oscillation period  P$_{osc}$=2195d. }
\end{figure*}

We find that our I-band magnitude time series displays a long-term increasing linear trend indicating that  the system is getting  brighter with time. 
Such a trend may be a segment of a very long activity cycle. Two significant (FAP $<$ 0.01\%) 
periodicities P$_1$ = 2017d (5.5\,yr)  and P$_2$ = 3563d (9.8\,yr) \rm are found superimposed to this linear trend.
Similar analysis was performed on the series of light curve brightest magnitudes, from which we derived a cycle   P$_1$ = 3763d (10.3\,yr), \rm and on the light curve amplitude,  
from which we derived a cycle P$_1$ = 2413d (6.6\,yr).  The periodogram analysis of the light curve faintest magnitudes did not reveal any significant periodicity. \rm Finally, we made the periodogram of the light curve phase of maxima and found an oscillation with period P$_{\rm osc}$ = 2195d (6\,yr). 
All these periods have a FAP $<$ 1\%.
The results are summarized in Table\,\ref{cycle}. In Fig.\,\ref{result}, we plot the complete time series of I-band magnitudes (top panel)  overplotted with the linear and sinusoidal fits
with the two periods (5.5 and 9.7\,yr). 
In the bottom panel, we plot the phases of maximum overplotted with  the 
sinusoidal fit with the period of 6\,yr. We note that our measurements allow us to sample three complete cycles of the phase variation which makes us rather confident that this periodicity may be real.
However, a much longer time series is needed to address this issue of possible long-term cycles.

\begin{figure*}
\begin{minipage}{18cm}
\centering
\includegraphics[scale = 0.5, trim = 0 0 0 0, clip, angle=0]{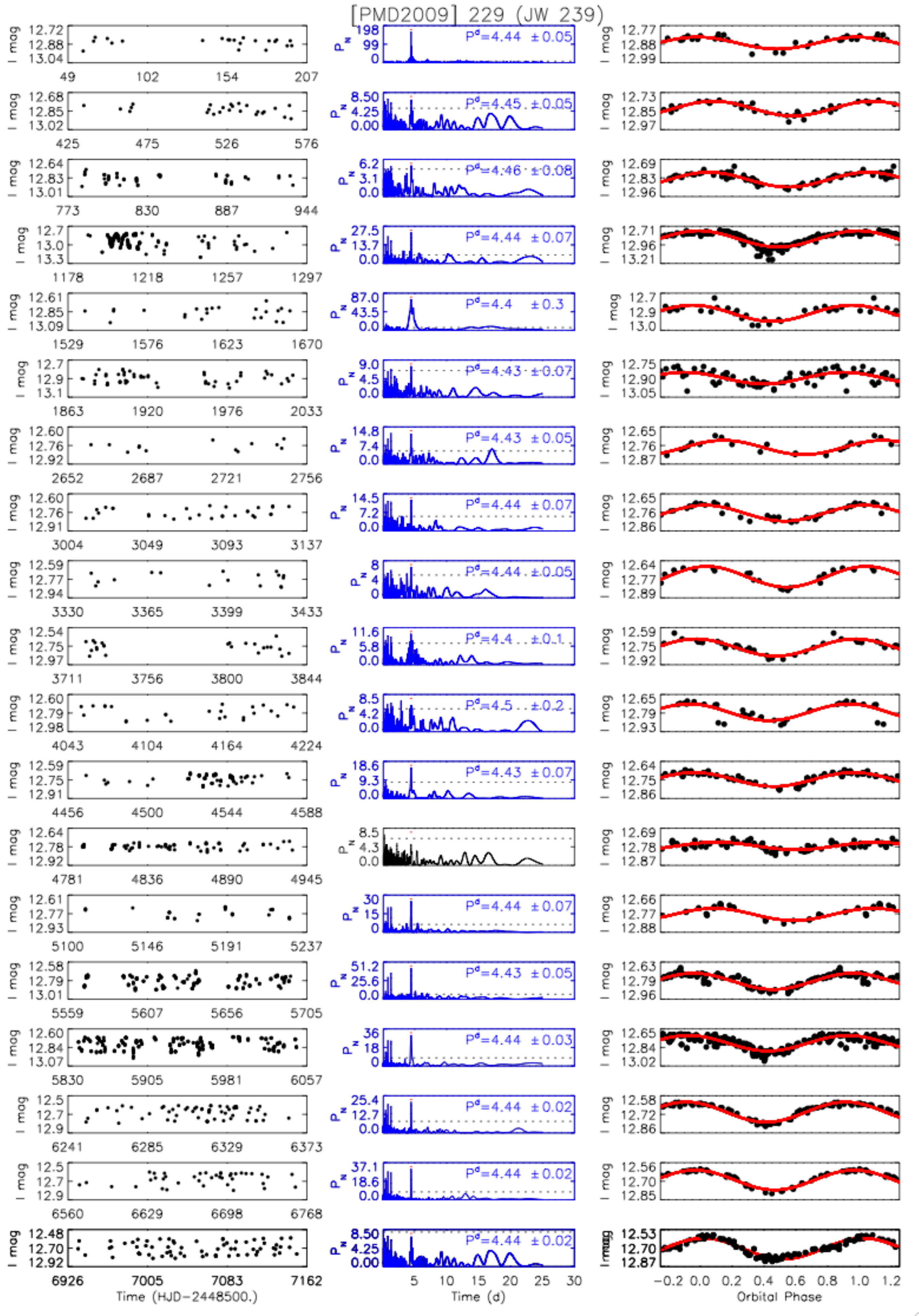}
\end{minipage}
\vspace{0cm}
\caption{\label{lightcurve} I-magnitude time series of V1481 Ori. \it Left \rm column displays the time segments of magnitudes versus HJD. 
\it Midlle \rm column displays the Lomb-Scargle periodograms with the 99\% confidence level indicated by the horizontal dashed line (black panel indicates no period detection above the 99\% confidence level). \it Right \rm column displays the light curves phased with the  ephemeris HJD = 2453658.8320  
+ 4.4351$\times$E. Solid lines represent the sinusoidal fit with the rotation period. }
\end{figure*}
\indent

\indent

\begin{table}
\caption{\label{data}Summary of I-band observations: mean epoch of each light curve, number of measurements, brightest I magnitude, light curve amplitude, and phase of light curve maximum.}
\begin{tabular}{|l|r|r|r|r|}
\hline
  \multicolumn{1}{|c|}{Mean Epoch} &
  \multicolumn{1}{c|}{N$_m$} &
  \multicolumn{1}{c|}{I$_{\rm min}$} &
  \multicolumn{1}{c|}{$\Delta$I} &
  \multicolumn{1}{c|}{Phase of} \\
   \multicolumn{1}{|c|}{(HJD)} &
  \multicolumn{1}{c|}{} &
  \multicolumn{1}{c|}{(mag)} &
  \multicolumn{1}{c|}{(mag)} &
  \multicolumn{1}{c|}{maximum} \\

\hline
 \multicolumn{5}{c|}{complete series} \\

  2452095.979 & 1522 & 12.725 & 0.148 & ...\\
   \hline
  2448628.717 & 22 & 12.831 & 0.082 & 0.822\\
  2449001.119 & 25 & 12.786 & 0.095 & 0.723\\
  2449359.189 & 58 & 12.788 & 0.104 & 0.752\\
  2449738.152 & 183 & 12.774 & 0.220 & 0.816\\
  2450100.182 & 23 & 12.784 & 0.167 & 0.854\\
  2450448.711 & 53 & 12.85 & 0.095 & 0.905\\
  2451204.652 & 13 & 12.731 & 0.063 & 0.659\\
  2451571.158 & 30 & 12.717 & 0.088 & 0.776\\
  2451882.279 & 17 & 12.677 & 0.147 & 0.763\\
  2452278.176 & 24 & 12.694 & 0.146 & 0.833\\
  2452634.223 & 26 & 12.72 & 0.124 & 0.875\\
  2453022.645 & 97 & 12.712 & 0.074 & 0.844\\
  2453363.492 & 77 & 12.759 & 0.038 & 0.704\\
  2453668.802 & 33 & 12.734 & 0.079 & 0.719\\
  2454132.187 & 210 & 12.73 & 0.144 & 0.847\\
  2454443.770 & 404 & 12.718 & 0.152 & 0.880\\
  2454807.325 & 65 & 12.63 & 0.147 & 0.892\\
  2455170.771 & 46 & 12.613 & 0.169 & 0.838\\
  2455544.276 & 69 & 12.613 & 0.189 & 0.783\\
\hline\end{tabular}
\end{table}

\begin{table}
\caption{\label{period}Rotation periods derived from periodogram analysis of the I-band time series segments.}
\begin{tabular}{|l|l|l|l|r|r|}
\hline
  \multicolumn{1}{c|}{HJD$_{initial}$} &
  \multicolumn{1}{c|}{HJD$_{final}$} &
  \multicolumn{1}{c|}{P(d)} &
  \multicolumn{1}{c|}{$\Delta$P} &
  \multicolumn{1}{c|}{Power} &
  \multicolumn{1}{c|}{Power}  \\
    \multicolumn{1}{c|}{} &
  \multicolumn{1}{c|}{} &
  \multicolumn{1}{c|}{} &
  \multicolumn{1}{c|}{} &
  \multicolumn{1}{c|}{} &
  \multicolumn{1}{c|}{@1\% FAP}  \\

\hline
2448539.858 & 2455267.122 & 4.43 & 0.05 & 164.85 & 8.23 \\
2448559.812 & 2448697.569 & 4.45 & 0.05 & 6.85 & 4.89 \\
2448925.745 & 2449072.542 & 4.46 & 0.07 & 5.16 & 4.58 \\
2449280.862 & 2449434.559 & 4.44 & 0.07 & 22.89 & 6.21 \\
2449688.734 & 2449789.552 & 4.43 & 0.30 & 72.53 & 6.47 \\
2450018.763 & 2450166.545 & 4.43 & 0.07 & 7.51 & 6.5 \\
2450369.842 & 2450523.580 & 4.43 & 0.05 & 11.96 & 5.22 \\
2451162.753 & 2451246.552 & 4.44 & 0.02 & 10.91 & 6.55 \\
2451514.687 & 2451629.562 & 4.44 & 0.07 & 12.06 & 5.69 \\
2451838.796 & 2451931.715 & 4.44 & 0.05 & 6.64 & 4.89 \\
2452221.728 & 2452338.596 & 4.42 & 0.15 & 9.67 & 6.71 \\
2452550.845 & 2452721.546 & 4.46 & 0.16 & 6.55 & 5.19 \\
2452960.757 & 2453079.620 & 4.43 & 0.07 & 15.47 & 8.08 \\
2453610.425 & 2453727.179 & 4.44 & 0.02 & 10.91 & 6.55 \\
2454069.261 & 2454197.081 & 4.44 & 0.07 & 25.03 & 6.32 \\
2454339.458 & 2454548.144 & 4.43 & 0.05 & 42.64 & 6.73 \\
2454751.446 & 2454863.204 & 4.44 & 0.03 & 29.88 & 7.91 \\
2455053.466 & 2455288.077 & 4.44 & 0.02 & 21.2 & 5.77  \\
2455436.451 & 2455652.100 & 4.44 & 0.02 & 30.91 & 6.55  \\
\hline\end{tabular}
\end{table}

\section{Orbital and Physical parameters}
\subsection{Radial velocity}
\begin{figure}
\begin{center}
\includegraphics[width=9.2cm]{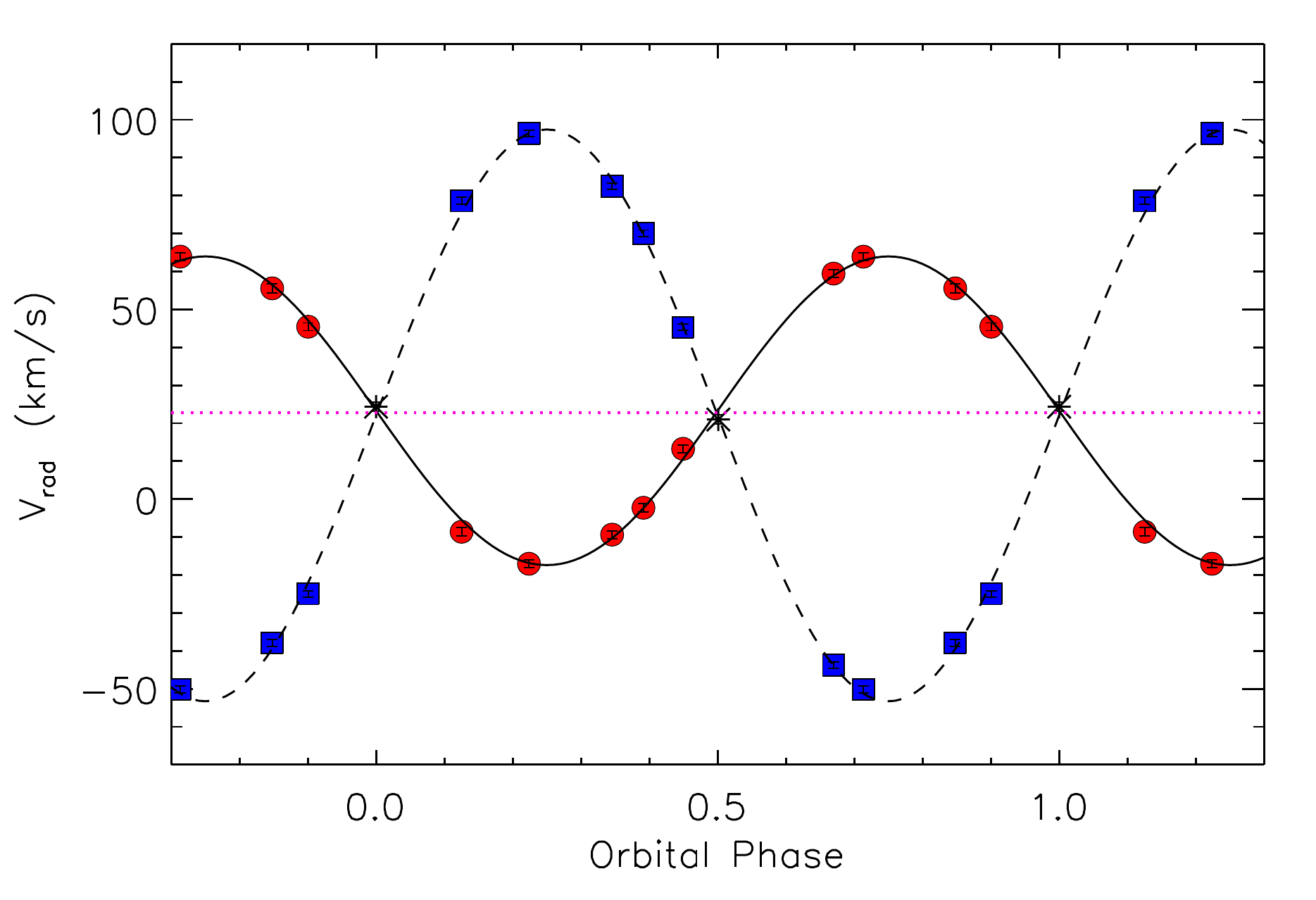}
\caption{\label{fig:radial_velocities}
Radial velocity curve of the binary system V1481\,Ori phased with the orbital period P = 4.433\,d. Circles and squares refer to the primary and the secondary components, respectively, while asterisks 
 are used when the two components were not resolved in the spectra. The RV errors are also shown and are smaller than the symbol size. The solid and dashed lines represent the 
 circular orbital solutions for the primary and secondary components, respectively, whereas the dotted line is the mean barycentric RV value.}
\end{center}
\end{figure}

We measured radial velocities (RVs) and their uncertainties following the procedure described in \cite{biazzoetal2009}, which is based on the analysis 
of the CCFs between stellar spectra and numerical masks (see that paper for details). In nine spectra, far 
from the conjunctions, we could resolve both binary components and get their respective RVs from two separate Gaussian fits. In the other 
two spectra, at phases very close to the conjunctions, we were able to see only one peak in the CCF and to measure  blended RVs. 
Regions containing telluric lines were removed. The RV values of both primary (more massive) and secondary components are listed in 
Table~\ref{tab:radial_velocities} and plotted in Fig.~\ref{fig:radial_velocities}, together with their errors. 
\begin{table}
\caption{Radial velocities of the binary system V1481\,Ori (1: primary component; 2: secondary component).}
\label{tab:radial_velocities}
\begin{center}  
\begin{tabular}{lrr}
\hline
\hline
\noalign{\smallskip}
$JD$           &  $V^{\rm A}_{\rm rad}$ &   $V^{\rm B}_{\rm rad}$ \\	 
 &   (km/s)	        &    (km/s)		  \\
\noalign{\smallskip}
\hline
2453658.8320$^\ast$      & $ 24.36\pm1.25$   	&		   \\
2453659.8226      & $-17.04\pm1.04$ &$ 96.34\pm0.83$   \\
2453660.8243      & $ 13.26\pm1.03$ &$ 45.27\pm0.82$   \\
2453661.8003      & $ 59.44\pm1.03$ &$-43.73\pm0.82$   \\
2453662.8240      & $ 45.43\pm1.02$ &$-24.96\pm0.81$   \\
2453663.8200      & $ -8.61\pm1.17$ &$ 78.60\pm0.94$   \\
2453664.7964      & $ -9.37\pm1.04$ &$ 82.46\pm0.83$   \\
2453678.7838$^\ast$      & $ 20.98\pm1.25$	&		   \\
2453679.7282      & $ 63.91\pm1.09$ &$-50.16\pm0.88$   \\
2453753.6614      & $ -2.29\pm1.10$ &$ 70.06\pm0.88$   \\
2453755.6823      & $ 55.55\pm1.14$ &$-37.94\pm0.91$   \\
\noalign{\smallskip}
\hline
\end{tabular}
\end{center}
Notes: $^\ast$ This phase is close to the conjunction, where only one CCF peak is visible.
\end{table}

\begin{figure}
\begin{minipage}{12cm}
\includegraphics[scale = 0.3, trim = 0 0 0 0, clip, angle=90]{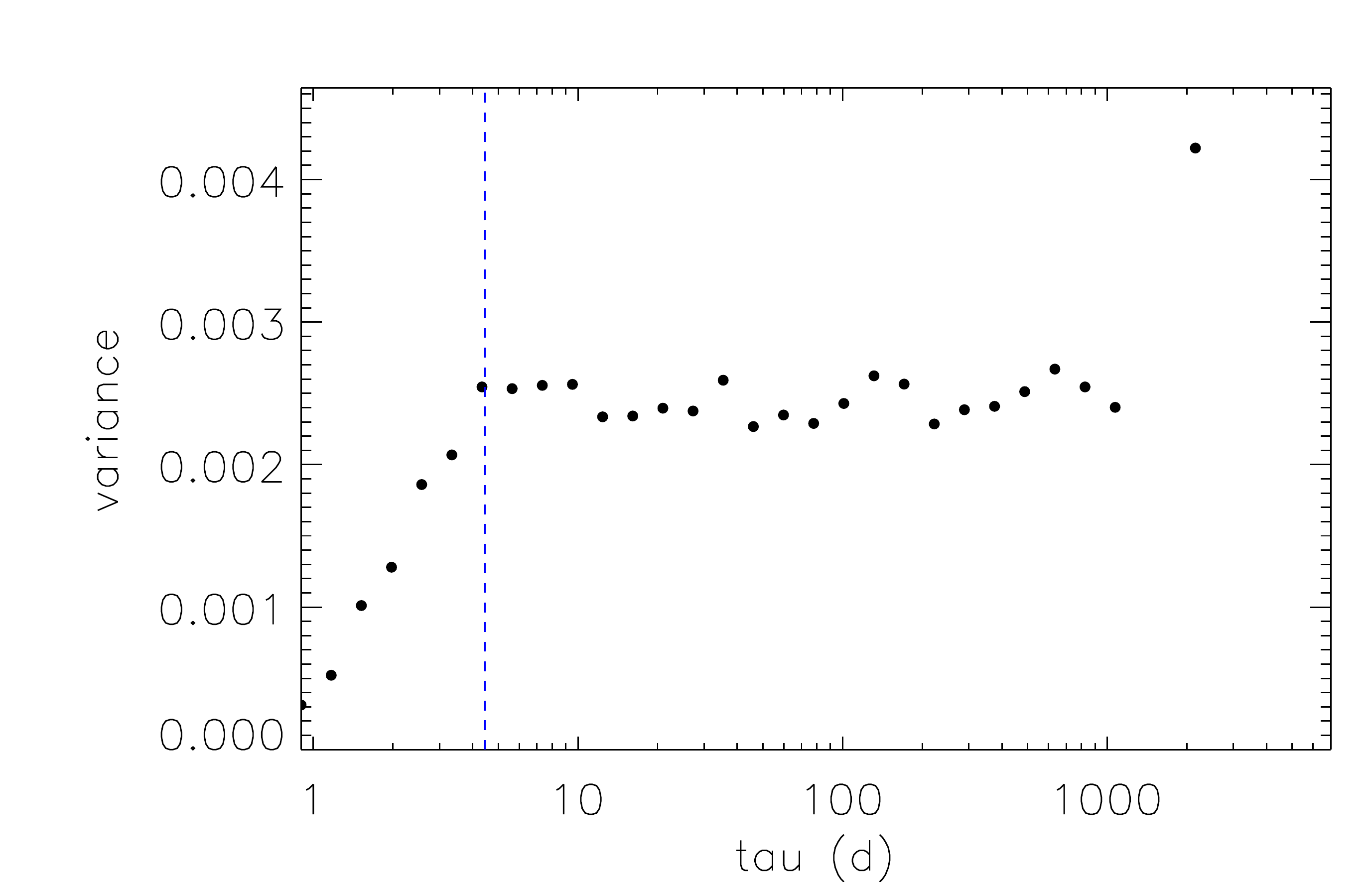}
\end{minipage}
\caption{\label{pool} Pooled variance profile of V1481 Ori. The vertical dashed line marks the rotation period.}
\end{figure}
\indent

We used the RV curves to search for orbital fitting parameters and their uncertainties, following the prescriptions of \cite{lucysweeney1971} for circular orbits and 
using the CURVEFIT routine (\citealt{Bevington69}) in IDL\footnote{IDL (Interactive Data Language) is a registered trademark of Exelis Visual Information 
Solutions.}. Thus, adopting the condition of null eccentricity ($e=0$), we obtained an orbital period $P_{\rm orb}=4.433$\,d, which is in agreement 
with the photometric one of 4.4351\,d (see Sect.\,4.2) and smaller than the cut-off value of 7.56 d found by \cite{meloetal2001} for orbital 
circularization in pre-main sequence binaries. With this orbital period and the initial heliocentric Julian Day (i.e. HJD0 = 2453658.8320; 
see  Sect.\,4.2) \rm we computed the orbital parameters of the binary system listed in Table\,\ref{orb_par}
(see also Fig.~\ref{fig:radial_velocities}).

\begin{table}
\caption{\label{orb_par}Measured orbital parameters: orbital period ($P_{\rm orb}$), barycentric velocity ($\gamma$), and RV semi-amplitude of the binary components ($k^{\rm A}$, $k^{\rm B}$).}
\begin{tabular}{lc}
\hline
$P_{\rm orb}$ (d) & 4.433$\pm$0.007\\
 $\gamma$  (km s$^{-1}$) & 22.7$\pm$0.3\\
 $k^{\rm A}$ (km s$^{-1}$ & $-$40.7$\pm$0.5 \\
  $k^{\rm B}$ (km s$^{-1}$ & 75.4$\pm$0.4 \\
  \hline

\end{tabular}
\end{table}
\subsection{Luminosity ratio}
\cite{Da Rio09} measured the following values of magnitudes and color: V = 15.153$\pm$0.004\,mag, I = 12.519$\pm$0.009\,mag, and V$-$I = 2.634\,mag. Their I magnitude
(obtained with the  ESO879 broadband filter) is not immediately comparable with our Bessel I magnitude. From TiO band they derive
an effective temperature T$_{\rm eff}$ = 3630\,K and a V-band extinction  A$_{\rm V}$ = 0.334 mag, and an accretion to total luminosity Log$_{\rm Ac}$ = $-$2.95 for the unresolved system.
They compare the HR position of V1481 Ori with different evolutionary models (\citealt{Da Rio10}), obtaining for the whole system  mass and age in the range from  0.29\,M$_\odot$ to 0.72\,M$_\odot$ and from 0.4\,Myr to 1.4\,Myr, using the \cite{D'Antona97} and \cite{Baraffe98} models, respectively.
 \cite{Da Rio12} subsequently obtained the intermediate
values M = 0.42\,M$_\odot$ and age = 1.1\,Myr (using \citealt{Siess00} models)  and M = 0.405\,M$_\odot$ and age = 0.75\,Myr (using  \citealt{Palla99} models). All these results were obtained before the binary nature of the system was known.
In contrast to earlier works, we are now in the position to obtain the parameters of the individual components.\\
  The CCF peaks show that the binary components have similar spectral types.
When spectral types of binary components are not very different (within
one spectral class), the relative intensities of the CCF peaks provide a good
proxy of individual stellar fluxes to the combined flux (see \citealt{Covino01},
and references therein). Since the stars are components of a binary system,
they share the same distance, and therefore the flux ratio provide a good approximation of
the luminosity ratio. We can therefore assume that the ratio of the CCF peaks
allows us to infer the luminosity ratio $L_{\rm B}/L_{\rm A}$ of the components. \rm
We find
a variable luminosity ratio with an average value of  0.79 between the two components and a mass ratio M$_{\rm B}$/M$_{\rm A}$ = 0.55.
Evolutionary models can help us to find masses and radii of both components. First, considering the hot spot as the cause of the observed variability, we will use the faintest and reddest
magnitude and color to position the system components on the Magnitude-T$_{\rm eff}$ diagram. Subsequently, we will consider the brightest magnitude and bluest color to position the components in the diagram. \\
We assume that the faintest observed I-band magnitude, I = 13.05\,mag, represents the unperturbed photosphere,  unaffected by either accretion effects
or starspots. Applying the extinction correction, A$_{\rm I}$ = 0.16\,mag, which is derived from  A$_{\rm V}$ = 0.33\,mag, assuming    A$_{\rm I}$/ A$_{\rm V}$ = 0.479 \rm  (\citealt{Cardelli89}),
and the distance d = 389.5$\pm$18\,pc (obtained from a weighted mean of the values given by \cite{Jeffries07}, \cite{Sandstrom07}, and   \cite{Bertout99}), we derive the absolute reddening-corrected magnitude I = 4.94$\pm$0.10 mag for the unresolved system. We obtain I = 4.49$\pm$0.10\,mag if we instead assume the brightest observed magnitude corresponds to the star immaculate state. 

From our spectroscopic and photometric analysis we know that the orbital and rotation periods are synchronized. Therefore,
it is reasonable to assume that both components have same inclination of their rotation axes. In such a case, the ratio
between the projected rotational velocities is equal to the ratio between the stellar radii, i.e., R$_{\rm B}$/R$_{\rm A}$ = 0.84$\pm$0.11.
We can use these observational constraints and the \cite{Baraffe98} models to find the masses of the individual components. Considering a possible age spread among the ONC members, we used models in the age range from 0.5 to 3 Myr. \rm 
We find that the measured mass and radius ratios are best reproduced by two stars with masses   
 M$_{\rm A}$ = 0.45\,M$_\odot$ and M$_{\rm B}$ = 0.25\,M$_\odot$ with radii  R$_{\rm A}$ = 1.97\,R$_\odot$ and R$_{\rm B}$ = 1.57\,R$_\odot$ at an age of 1\,Myr and with a luminosity ratio L$_{\rm B}$/L$_{\rm A}$ = 0.54.

\begin{figure}
\begin{minipage}{10cm}
\includegraphics[scale = 0.30, trim = 0 0 0 0, clip, angle=90]{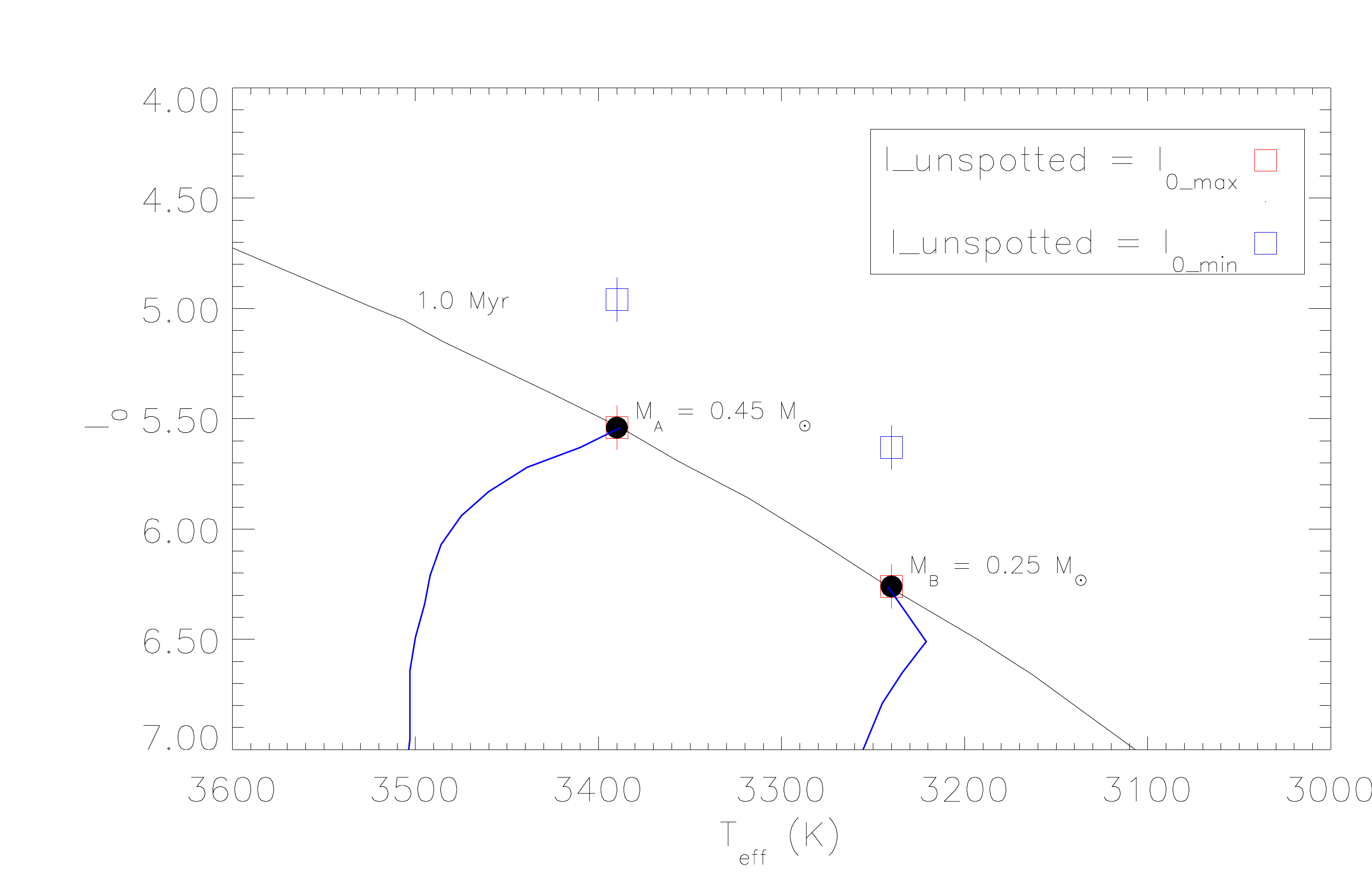}
\end{minipage}
\vspace{-0cm}
\caption{\label{hr} I$_0$ vs. Teff diagram of V1481 Ori. Black and blue solid lines represent theoretical isochrones and mass evolutionary tracks, respectively, from {Baraffe et al. (1998)}. Black bullets represent the magnitudes expected from a model at 1 Myr. Red and blue open squares represent the observed magnitudes in the hypothesis that the unspotted magnitude corresponds to the faintest and brightest observed values, respectively. }
\end{figure}

The model-derived absolute I-band magnitudes and effective temperatures
for the two components are I$_{\rm A}$ = 5.54\,mag and I$_{\rm B}$ = 6.26\,mag and
T$_{\rm A}$ = 3390\,K and T$_{\rm B}$ = 3240\,K (black bullets in Fig.\,\ref{hr}).  Since the two components are at the same distance from the Sun and have similar spectral types, we assume that the flux ratio
F$_{\rm B}$/F$_{\rm A}$ $\simeq$  L$_{\rm B}$/L$_{\rm A}$ = 0.54 and, therefore, we can infer the magnitude correction to apply to the observed absolute magnitude (I = 4.94$\pm$0.10 mag) to compute the absolute magnitudes of the individual components.
Specifically, we obtain I$_{\rm A}$ = 5.41$\pm$10 mag and I$_{\rm B}$ =  6.08$\pm$0.10 mag (red squares in Fig.\,\ref{hr}).  
In contrast to the previous assumption, if we consider the brightest observed magnitude I = 12.60\,mag as corresponding to the immaculate state, 
we obtain I$_{\rm A}$ = 4.95$\pm$0.10\,mag and I$_{\rm B}$ =  5.61$\pm$0.10\,mag (blue squares in Fig.\,\ref{hr}), which are irreconcilable with the model values.

The magnitude difference $\Delta$I = 0.45$\pm$0.02 mag between the faintest and brightest state implies that
the observed flux, and consequently the luminosity at the brightest level, is about a factor 1.5 larger than expected from the model. This is in very good agreement with the mean luminosity ratio L$_{\rm B}$/L$_{\rm A}$ = 0.79 derived from
our spectroscopic analysis, which is a factor 1.5 larger than the value L$_{\rm B}$/L$_{\rm A}$ = 0.54 derived from the model.
This result is the first clear evidence that the mentioned hot spot is on the secondary component
of the system.\\
Combining stellar radii, projected rotational velocities $v\sin{i}_A$ = 19.2$\pm$1.1 kms$^{-1}$ and $v\sin{i}_B$ = 16.1$\pm$0.9 kms$^{-1}$ measured by \cite{biazzoetal2009}, and our rotation period we find the inclination of the stellar rotation axes to be $i$$_{\rm A}$ $\simeq$ $i$$_{\rm B}$ = 60$^{\circ}$.

Considering that the separation between the components is about 10 solar radii  as inferred from the Kepler law (and smaller from surface to surface),
which is the typical size of magnetospheres around TTS stars, and that these magnetospheres are likely interacting,
it is unlikely that each component has its own freely orbiting 
 circumstellar disc. Rather the system presumably possesses a circumbinary disc that falls primarily onto the secondary, because its orbit carries it a bit closer to the inner edge of any circumbinary disc with respect to the primary component.
Consequently, the secondary component exhibits a hot spot that is responsible
for the observed variability of the whole system and makes the component B up to one and a one-half times more luminous
than expected from a quiet state.

\subsection{Hot spot modeling}
To further investigate  our hypothesis of a hot spot, we modeled the observed V and I magnitude variations, by selecting the light curve (\#14) obtained at the mean epoch HJD 2453668.802 almost simultaneously with the spectroscopic observations.
We used the \cite{Prsa05} PHOEBE program to simultaneously fit the V and I light curves, as well as the spectroscopic RV curves. Initial guess values of the spot parameters were obtained  after preliminary models of the individual light curves using Binary Maker 3.0 (\citealt{Bradstreet04}).  We used as input values the known orbital and physical parameters presented in this Section.
The gravity-darkening coefficient has been assumed to be $\nu$ = 0.25 (\citealt{Kopal59}), and  the limb-darkening coefficients from {Claret  et al. (1995) were  adopted. Thanks to the dependence of the light curve amplitude on the photometric band wavelength, we could constrain the spot temperature contrast and better determine the area of the spots  responsible for the flux rotational modulation. 
An inspection by eye  in the top panel of Fig.\,\ref{3Dmodel} \rm immediately reveals that the two light curves do not have the same shape, so it is not a surprise that fitting both equally well failed.  On the other hand, the fit to the V light curve is satisfactory, in that the models fit all the bumps and wiggles of that curve.  However, the same spot in I-band flux fails,
although it predicts the correct amplitude. We also could successfully  fit  only the I-band flux curve (reduced $\chi_{V}^{2}$ = 0.70 chi square). However, when we use same parameters for the V light curve, they do not  fit at all (reduced $\chi_{I}^{2}$ = 2.4).  
We have to remind the reader that we are adopting a very simplified scenario, where the observed variability is assumed to be caused by only the hot spots on the secondary component  (see bottom panel of Fig.\,\ref{3Dmodel}). \rm However, since both components are magnetically active, we expect that also cool spots and hot faculae on both components can contribute to the observed variability, revealing the limits of our modeling approach.

\begin{figure}
\begin{minipage}{10cm}
\includegraphics[scale = 0.3, trim = 0 0 0 0, clip, angle=90]{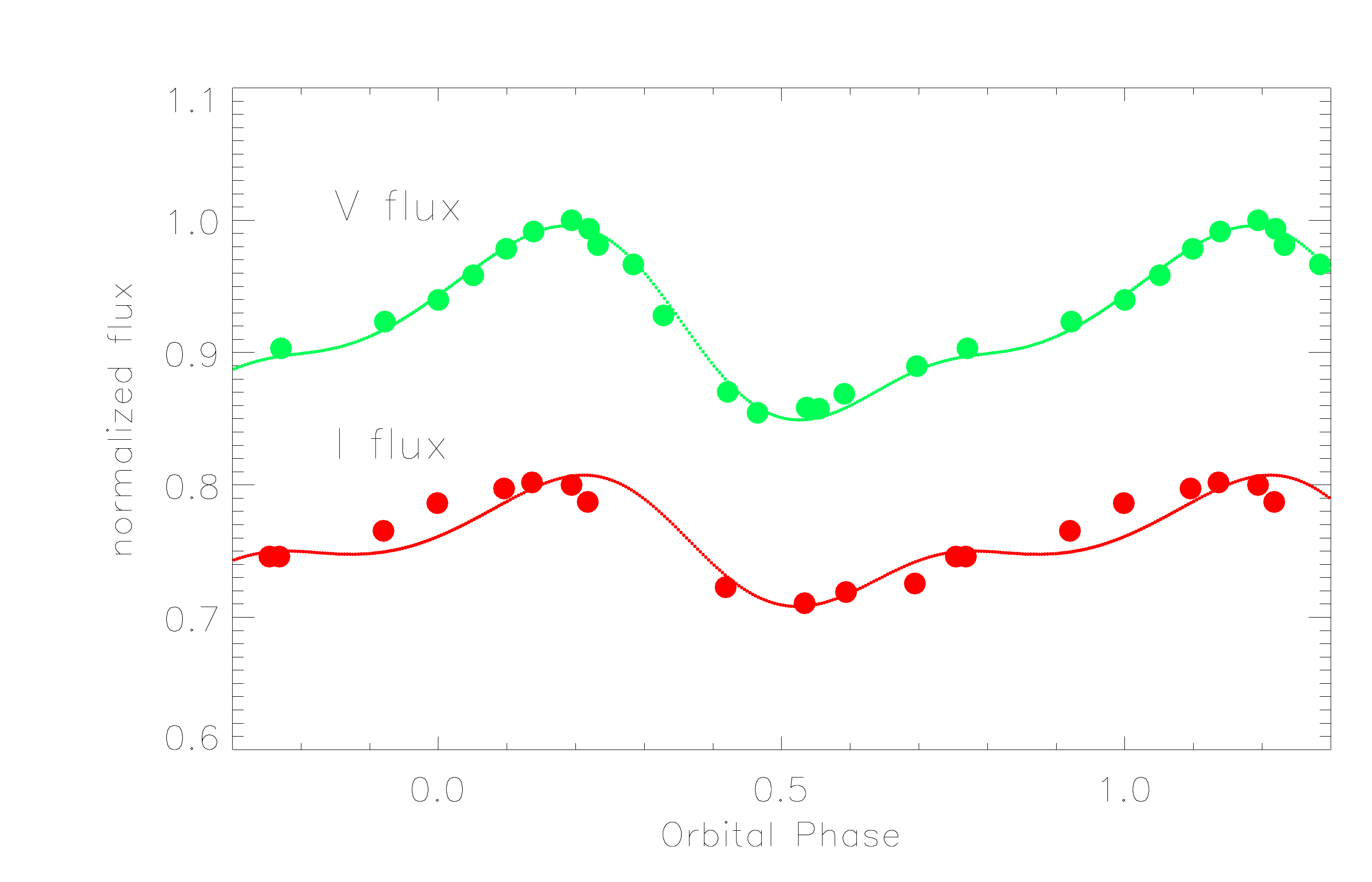}\\
\includegraphics[scale = 0.4, trim = 0 0 0 0, clip, angle=0]{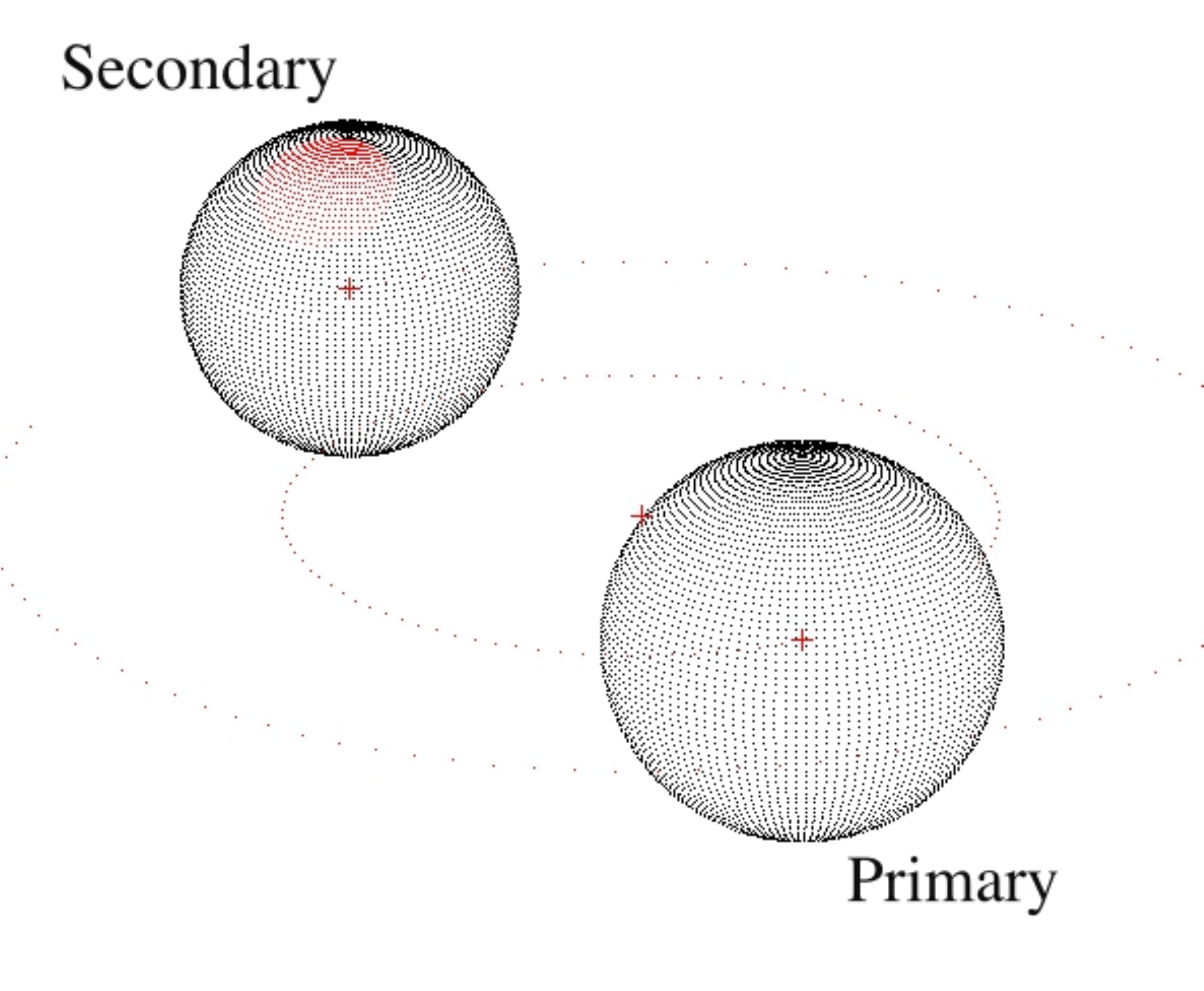}
\end{minipage}
\vspace{-0cm}
\caption{\label{3Dmodel} \it top \rm Normalized V- and I-fluxes (arbitrarily shifted) versus rotation phase at the mean epoch HJD 2453668.802 with overplotted the fit obtained from the hot spot modelling
on the secondary component. \it bottom \rm Pictorial 3D representation of V1148 Ori with the hot spot on the secondary component. The dotted lines represent the orbital paths of the system components.}
\end{figure}

The best fit of both curves ($\chi_{V}^{2}$$ = 0.70$ and $\chi_{I}^{2}$$ = 2.4$) is achieved assuming
on the secondary B component a circular spot with radius r = 20$^{\circ}$ and a temperature contrast T$_{spot}$/T$_{eff}$ = 1.13
($\Delta$T = 420\,K), whose best visibility is at phase $\phi$ = 0.13 (see Fig.\,\ref{3Dmodel}). 
Assuming that the excess flux emitted by the hot spot comes from the accretion luminosity, we estimate a lower limit to the mass accrection rate of 9.3$\times$10$^{-10}$\,M$_\odot$\,yr$^{-1}$,  not implausible for single late-type T Tau stars of similar mass (e.g., \citealt{Ingleby13}).

In Fig.\,\ref{synoptic}, we plot, from top to bottom, the radial velocity curves, the L$_B$/L$_A$ luminosity ratio, and the V-band flux curve, which is normalized to the brightest light curve magnitude, versus the orbital phase. We note a very good correspondence between the luminosity ratio and the V light curves. When the flux is maximum, i.e. when the hot spot has best visibility, also the luminosity ratio is maximum. This correspondence supports the presence of the hot spot on the secondary component. The two maxima are at about orbital phase $\phi$ = 0.19, which means that the hot spot photocenter is offset by about 70$^{\circ}$ in longitude from the conjunction of the two components.\\

 Our spot modeling is focussed only on the 2005  observation season during which we could also get the spectra. However, an inspection at the top panel of Fig.\,\ref{result} reveals that  brightest, faintest magnitudes, and light curve amplitude all vary from season to season. Such variations likely correspond to different combinations of  area and average latitude of the accretion  hot spot. As well, in contrast to the 2005 season, when the variability was dominated by the hot spot, also cool spots may contribute in other seasons to the observed variations. \\ \rm  

\begin{figure}
\begin{minipage}{10cm}
\includegraphics[scale = 0.45, trim = 0 0 0 0, clip, angle=0]{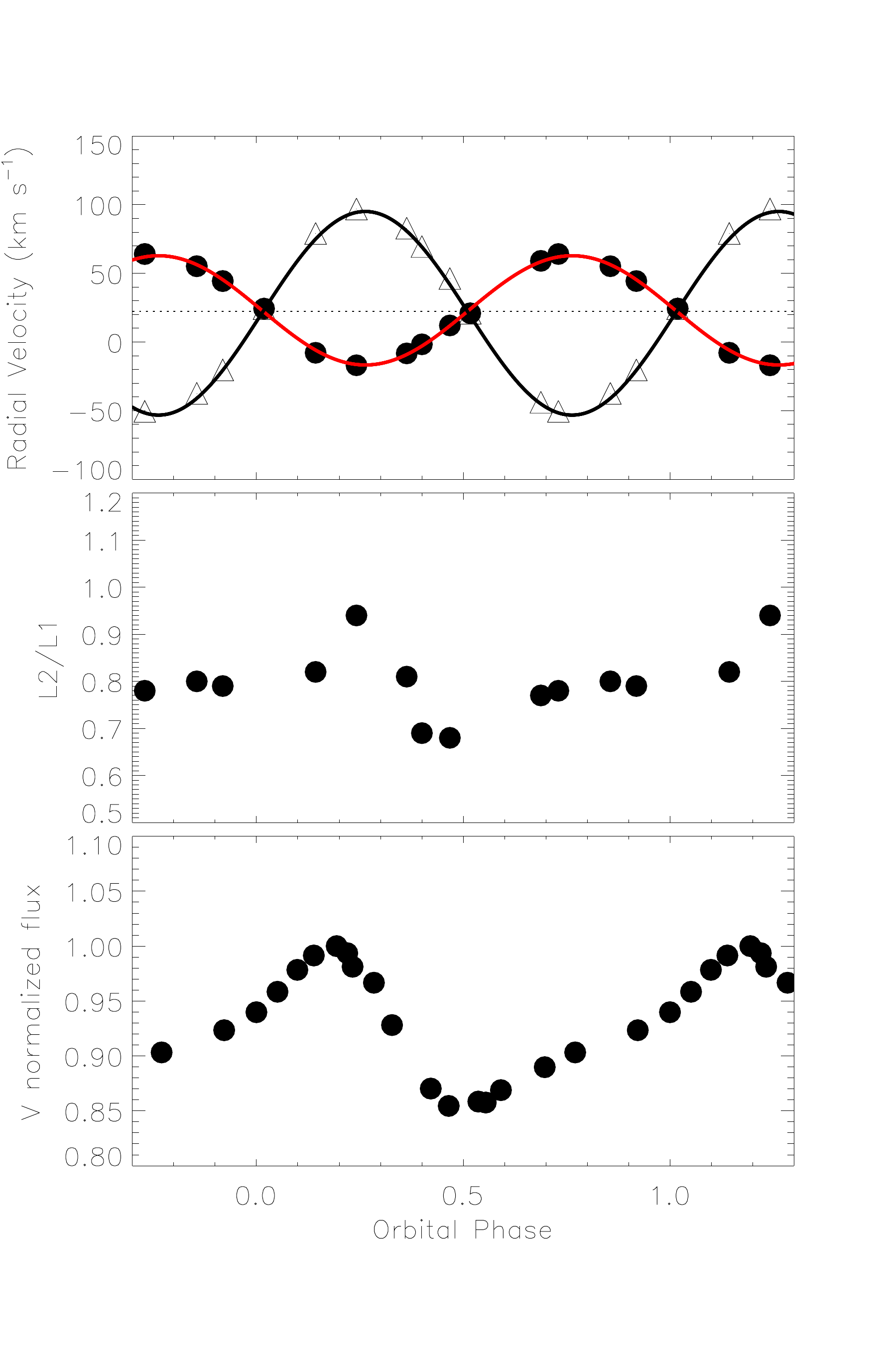}
\end{minipage}
\vspace{-0cm}
\caption{\label{synoptic} From top to bottom, RV curves, luminosity ratio curve, and V-band flux (normalized to the brightest magnitude) versus orbital phase.}
\end{figure}

\section{Discussion}
The available data show that the components of V1481 Ori seem to belong to Class II (CTTS). In fact,  the  H$\alpha$ line, the prominent  IR excess, and the presence of an accreting hot spot  as inferred from our modeling, are  consistent with the existence of a circumbinary accretion disc typical of CTTS stars. Moreover,  its optical light curve amplitude, after correction for the light dilution effect, is  $\Delta$I $>$ 0.44 mag ($\Delta$V $>$ 0.70 mag), which is a value observed in CTTS stars, whereas WTTS stars generally exhibit lower amplitudes (see, e.g. {Herbst et al. 1994). Despite the stochastic variability generally associated with accretion phenomena, V1481 Ori is one of the known ONC members with the most stable light curves (see Fig.\,\ref{lightcurve}). 
The pooled variance profile shown in Fig.\,\ref{pool} also shows that the variability mostly comes from
the rotation and additional variability sources start contributing at much longer time scales  ($>$ 3 yr). 
This behavior suggests that the secondary component of V1481 Ori has an accretion disc though the accretion is moderate and quite stable; or, that the accretion from a
 circumbinary disc is primarily onto the secondary, as models suggest.\\
Our measurements of the luminosity ratio show that it is variable in the range from 0.68 to 0.94. This variation is correlated to the orbital phase and can be accounted by a hot spot covering about 3.5\% of the photosphere. However, the minimum value 0.69 is about 30\% larger than expected in the case of no accretion. This means that we must invoke the presence of an additional component of bright regions  that are quite visible at all phases and therefore not undergoing rotational modulation. In this case, the shape of the hot spot must be quite more complicated than that we could model. We can image that the infalling gas from the disc
produces a quite uniform bright belt at the latitude where the hot spot is located, with superimposed the hot spot.\\
The existence of a circumbinary disc that is accreting material on the secondary component gives a strong observational support to the models of \cite{Artymowicz96}. These models predict that, in the case of close T Tauri binaries with low eccentricity and with unequal component masses, the mass flow from the circumbinary disc occurs preferentially onto the lower mass object. Our findings, on the contrary, give less support to other models, e.g., by 
\cite{de Val-Borro11}, that predict the accretion to occur on both components, although we cannot rule out the presence of at least some accretion onto the primary.\\
Fully convective stars such as V1481 Ori are expected to rotate as a solid body. Indeed, we do not
detect any evidence of SDR from seasonal rotation period variations at the 0.7\% level, which is the 3$\sigma$ precision of our rotation period measurements. On the other hand, young main sequence stars with a rotation period of about 5 days generally exhibit  variations of the measured photometric rotation period  $\Delta$P/P  up to  5\%, which are attributed to surface differential rotation \rm (e.g., \citealt{Messina03}). What we unexpectedly found is a possible slow oscillation of the phase of maximum with period of about 6 years. Such an oscillation can arise from a variation of the angular velocity with which the hot spot producing the light modulation is carried out across the photosphere. This angular velocity exhibits a variation $\Delta \Omega / \Omega$ = 0.065\%, which is a factor 20 smaller than the precision with which our measurements can reveal rotation period variations. This may be a nice example case of how  powerful the analysis of phase migration of light curve minima in SDR studies can be.
The angular velocity variation we observed can be due to a periodic latitude migration of the hot spot on a differentially rotating star. In this case, we measure a lower limit to the amplitude of SDR,
since the latitude interval spanned by the hot spot is limited. 

Because our star is probably locked to its accretion disc by its strong magnetic field, an alternative explanation is that the disc and the star exchange angular momentum back and forth in a cyclic fashion. In this case, assuming the stellar parameters given above and an internal structure as a polytrope of index n = 3/2, we can estimate the minimum magnetic field strength required to transfer a sufficient amount of angular momentum during an oscillation period to account for the observed amplitude  (see Appendix A). \rm We find a minimum field strength  of about 650\,G at the stellar surface, that is in agreement with the strong surface fields found in T Tau stars.

\section{Conclusions}
We have carried out a photometric and spectroscopic study of the  SB2 spectroscopic binary V1481 Ori, a member of the Orion Nebula Cluster.
Spectroscopic data were collected at the ESO/VLT and  photometric data were collected at the Indian Astronomical Observatory since 2004. The latter were 
complemented with photometric observations performed at Van Vleck Observatory since 1991, totaling a series of about 20 yr of I-band photometry.
Our spectroscopic analysis has allowed us to confirm the binary nature of this system, obtaining the radial velocity curves of both components, 
from which we derived the orbital period P$_{\rm orb}$ = 4.433\,d, the mass ratio M$_{\rm B}$/M$_{\rm A}$ = 0.54, 
and  the luminosity ratio  L$_{\rm B}$/L$_{\rm A}$, which is found to vary with orbital phase.
A comparison with evolutionary models from Baraffe et al. (1998) allowed us to infer  masses  M$_{\rm A}$ = 0.45\,M$_\odot$ and M$_{\rm B}$ = 0.25\,M$_\odot$ and
 radii  R$_{\rm A}$ = 1.97\,R$_\odot$ and R$_{\rm B}$ = 1.57\,R$_\odot$ of both components, and the inclination of the rotation axes $i$$_{\rm A}$ $\simeq$ $i$$_{\rm B}$ = 60$^{\circ}$.

The analysis of photometric data has allowed us to detect a very stable light modulation with an average period P$_{\rm rot}$ = 4.4351\,d that we
attribute to the presence of a hot spot, which is presumably generated by material accreting from a circumbinary disc onto the lower-mass component, and is carried in and 
out of view during the orbital revolution.
Although both components are expected to host some level of magnetic activity (starspots) due to the deep convection zone and the fast rotation, 
the observed variability arising from this hot spot seems to be dominant with respect to any additional contribution from cool spots.
The presence of a hot spot  on the secondary component is also found by our modeling of the multi-band light curves and RV curves collected in the 2005
season. 

We find that luminosity ratio variations correlate very well with the I-band flux variations, in the sense that the maximum luminosity ratio is observed at the same phases (0.2-0.3)
that the light curve has its maximum. This circumstance gives strong support to the presence of a hot spot as cause of the observed variability. The hot spot is located at about 70 deg 
from the substellar longitude. Our findings favor those accretion models predicting, in the case of close T Tauri binaries with low eccentricity and with unequal component masses, 
the mass flow from the circumbinary disc to occur preferentially onto the lower mass object.

 From the migration in phase of the light curve maximum we infer a variation of the photometric period of 0.065\%. That is very close to rigid-body rotation, as expected by the models 
 mentioned in the Introduction.  We find an interesting periodic oscillation of about 6 yr that can arise from a variation of the angular velocity with which the hot spot producing the light modulation is carried out across 
 the photosphere. The angular velocity variation we observed can be due to a periodic latitude migration of the hot spot on a differentially rotating star.
 An alternative explanation is that the disc and the star exchange angular momentum back and forth in a cyclic fashion.
 In this case, we find a minimum required field strength of about 650\,G at the stellar surface, that is in agreement with the strong surface fields found in T Tau stars.

\begin{table}
\caption{\label{cycle}Summary of search for long-term periodic variations: activity indicator, primary/secondary cycle, length of cycle (d), power of the highest periodogram peak, power corresponding to a 99\% significance level.}
\begin{tabular}{|l|r|r|r|r|}
\hline
  \multicolumn{1}{|c|}{Activity} &
  \multicolumn{1}{c|}{Cycle} &
  \multicolumn{1}{c|}{P$\pm$$\Delta$P} &
  \multicolumn{1}{c|}{P$_N$} &
  \multicolumn{1}{c|}{P$_{N99\%}$} \\
   \multicolumn{1}{|c|}{Indicator} &
  \multicolumn{1}{c|}{} &
  \multicolumn{1}{c|}{(d)} &
  \multicolumn{1}{c|}{} &
  \multicolumn{1}{c|}{} \\

\hline

I$_{\rm mag}$ &	P$_1$	&2017$\pm$141& 	99&	7\\
		&P$_2$	&3563$\pm$440	&97	&7\\
		
I $_{\rm min}$	&	P$_1$	&3763$\pm$76&	2.68	&1.43\\
I $_{\rm ampl}$&	P$_1$	&2413$\pm$31&	6.27 &4.50\\

Phase of maximum&	P$_1$	&2195$\pm$26	&6.5	&3.16\\
\hline\end{tabular}
\end{table}

\appendix 

\section{Magnetic coupling between a star and its disc: angular momentum exchange}
\label{AppendixA}

The variation of the spin angular momentum of the star $L$ is given by: 
\begin{equation}
\frac{dL}{dt} = \tau,
\label{eq1}
\end{equation}
where $t$ is the time and $\tau$ the torque due to the Maxwell stresses produced by the magnetic field that couples the star with its disc. The torque applied at the surface of the star is given by the integral of the Maxwell stress $B_{r} B_{\phi}$ applied there \citep[cf.][]{Lanza07}:
\begin{equation}
\tau = -\frac{1}{\mu} \int_{S} R \sin \theta B_{r} B_{\phi} \, dS,
\end{equation}
where $\mu$ is the magnetic permeability of the plasma, $R$ the radius of the star, $\theta$ the colatitude measured from the North pole, $B_{r}$ the radial and $B_{\phi}$ the azimuthal component of the magnetic field at the surface of the star, and $S$ the stellar surface. 
Assuming that the field components are axisymmetric, we obtain:
\begin{equation}
\tau = -\frac{\pi^{2}}{\mu} R^{3} \langle B_{r} B_{\phi} \rangle,
\label{eq3}
\end{equation}
where the brackets indicate the mean value of the Maxwell stress over the surface of the star. If the magnetic stresses modulate the angular velocity $\Omega$ of the star with a period $P_{\rm mod}$, the amplitude of the variation of the angular momentum is given by:
\begin{equation}
\frac{dL}{dt} = \frac{2\pi I}{P_{\rm mod}} \Delta \Omega,
\label{eq4}
\end{equation}
where $I$ is the moment of inertia of the star and $\Delta \Omega$ the amplitude of its angular velocity modulation assumed to oscillate sinusoidally as a function of the time. The moment of inertia of the star can be written as $I = kMR^{2}$, where $k = 0.205$ for a polytrope of index $n=3/2$, is a constant depending on the density stratification inside the star and $M$ is the mass of the star. Considering Equations (\ref{eq1}), we can equate the right-hand sides of  (\ref{eq3}) and (\ref{eq4}) and with some re-arrangement of the terms we obtain:
\begin{equation}
\langle B_{r} B_{\phi} \rangle = \frac{2 k  \mu}{\pi R P_{\rm mod}} M \Delta \Omega,
\end{equation}
that we use to estimate the Maxwell stress at the surface of the star. The minimum magnetic field strength $B_{\rm min}$ corresponds to the case when $B_{r} = B_{\phi}$ and is given by: $B_{\rm min} = (\langle B_{r} B_{\phi} \rangle)^{1/2}$.

\clearpage
\end{document}